\documentclass[11pt,twoside,
]{article}

\usepackage{baaa2009}
\usepackage{graphicx}
\usepackage{subfigure}
\usepackage{psfrag}
\usepackage{amssymb}
\usepackage[spanish,activeacute,english]{babel}
\usepackage[latin1]{inputenc}
\usepackage[T1]{fontenc} % Computer Modern (CM) fonts
\usepackage{ae,aecompl} % and: dvips -Pcmz -Pamz macros_aaa.dvi
\usepackage{latexsym}
\usepackage{verbatim}
\usepackage{amsmath}
\usepackage{stmaryrd}
\usepackage{amsfonts}
\usepackage{amssymb}
\usepackage{wasysym}
\usepackage[colorlinks=true,dvips]{hyperref}
\begin{document}
%%% SELECCIONE EL IDIOMA EN QUE SE ESCRIBE EL ARTÍCULO:              %%%%
%\myselectspanish
\myselectenglish
%%%%%%%%%%%%%%%%%%%%%%%%%%%%%%%%%%%%%%%%%%%%%%%%%%%%%%%%%%%%%%%%%%%%%%%%%%
\vskip 1.0cm \markboth{ A. Etchegoyen \textit{for the Pierre Auger
Collaboration}} {Science and Detectors of the Pierre Auger
Observatory}

\pagestyle{myheadings}
%%%% DESCOMENTE LA LINEA QUE DESCRIBE EL CARACTER DE SU TRABAJO       %%%%
\vspace*{0.5cm}
\noindent INVITED TALK
%\noindent PRESENTACIÓN ORAL
%\noindent PRESENTACIÓN MURAL
%\noindent RESUMEN
\vskip 0.3cm
\title{Science and Detectors of the Pierre Auger Observatory}

%\title{ Template paper for publication in the Bulletin of the
%Argentinian Astronomical Association with instructions for the use of
%\LaTeX{}}

\author{A. Etchegoyen $^{1,2}$ \textit{for the Pierre Auger Collaboration}}

\affil{%
  (1) ITeDA,Instituto de Tecnolog\'ias en Detecci\'on y Astropart\'iculas (CNEA, CONICET, UNSAM)\\
  (2) UTN-FRBA \\
}

\begin{abstract} The high energy spectrum of cosmic rays presents three distinct
traits, the second knee, the ankle, and the GZK cutoff and as
such, a thorough understanding of cosmic rays encompasses the
study of these three features. It is in the second knee - ankle
region where cosmic ray sources change from a galactic origin to
an extragalactic one. At the higher cutoff energies, the arrival
directions show an anisotropy related to the near extragalactic
sky. The Pierre Auger Observatory is currently designed to help to
unravel these features by performing both spectrum and composition
measurements with unprecedented accuracy. The primary particle
type in the second knee - ankle region will be studied both with
fluorescence telescopes and muon counters giving the air shower
longitudinal profiles and muon contents, respectively.
\end{abstract}

\begin{resumen}
 El espectro de rayos c\'osmicos de altas energ\'ias presenta tres
 rasgos distintivos, la segunda rodilla, el tobillo y el corte
 GZK y por ende comprender en detalle a estos rayos c\'osmicos
 implica el estudio de estas tres caracter\'isticas. Es en la zona
 de la segunda rodilla - tobillo en donde acaece el cambio de fuentes de
 rayos c\'osmicos de gal\'acticas a extra gal\'acticas. Por otro lado, en
 la zona de las energ\'ias  m\'as altas se detecta anisotrop\'ia en
 la direcci\'on de arribo relacionada con el espacio
 extragal\'actico cercano. El Observatorio Pierre Auger est\'a
 dise\~nado para el estudio de estos tres rasgos realizando tanto
 estudios del espectro como de la composici\'on qu\'imica de los
 rayos c\'osmicos primarios con una precisi\'on sin precedentes. La
 composici\'on qu\'imica del primario ser\'a estudiada con telescopios
 de fluorescencia y con contadores de muones obteni\'endose as\'i
 los perfiles longitudinales y los contenidos mu\'onicos de los chub\'ascos
 c\'osmicos.
\end{resumen}

\section{Introduction}

The Pierre Auger Observatory was built to detect the highest
energy cosmic rays known in nature with two distinctive design
features, a large size and a hybrid detection system in an effort
to detect a large number of events per year with minimum
systematic uncertainties. The southern component of the Auger
Observatory is located in the west of Argentina, in the Province
of Mendoza where it spans an area of 3000 km$^{2}$ covered with
1600 water Cherenkov detectors (SD, surface detectors) deployed on
a 1500 m triangular grid with four buildings on the array
periphery lodging six fluorescence telescope systems (FD,
fluorescence detectors) each one with a $30^{\circ }\times
30^{\circ}$ elevation and azimuth field of view (Abraham et al.
(2004, 2008a)). With such a geometry, the Observatory has been
able to cast light on two spectral features (Abraham et al.
(2009a)) at the highest energies, the ankle and the GZK-cutoff
(Greisen (1966), Zatsepin et al. (1966)).

The ankle refers to a break in the spectrum which is found at $
\sim 10^{18.6}$ eV. by the Auger Observatory. The GZK-cutoff
(named after Greisen, Zatsepin, and Kuz'min who suggested it) is a
suppression of the cosmic ray flux at very high energies which is
reported at $\sim 10^{19.5}$ eV (Abraham et al. (2009a)). This
suppression is pertinent to the reported anisotropy in the arrival
directions of cosmic rays with energies above $\sim 6 \times
10^{19}$ eV (Abraham et al. (2008b)). This anisotropy may open a
new window in astronomy research, charged-particle source
identification, since the very high primary-particle energies
might permit to trace back the arrival directions and find the
cosmic sources. At lower energies the electromagnetic fields
deflect charged-particle trajectories rendering impossible to
identify the sources, but still composition studies should help to
discriminate whether the sources are galactic or extragalactic and
the energy region where the transition occurs. Also, non-charged
particles (e.g. neutrons and photons) might produce detectable
point-like galactic anisotropies. Neutrons with energies above
$\sim$ 1.0 EeV could arrive from the galactic-center region
without decaying. Photon-induced anisotropies would be easier to
identify with the Auger Observatory enhancements (see below) for
instance with muon counters, because photon showers have a
vanishing muon content.

\begin{figure}[!h]
\begin{center}
\includegraphics[width= 7.0 cm]{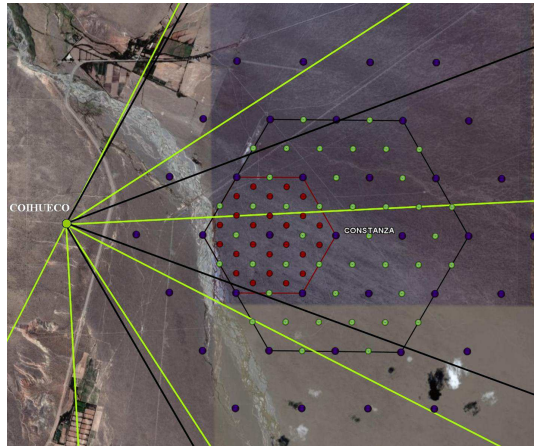}~\hfill
\caption{Auger enhancements layout. Green lines limit the
$0^{\circ }-30^{\circ }\times 0^{\circ }-30^{\circ}$  elevation
and azimuth FOV of the original 6 telescopes on Cerro Coihueco and
the black lines the $30^{\circ }-60^{\circ }\times 0^{\circ
}-30^{\circ}$ FOV for the 3 HEAT telescopes. The two hexagons
limit the AMIGA infilled areas of 5.9 and 23.5 km$^{2}$ with 433
and 750 m triangular grid detector spacings, respectively. Each
dot within these hexagons represents a pair of a water Cherenkov
detector and a muon counter. The center dot, named Constanza, is
placed $\sim$ 6.0 away from Cerro Coihueco. \label{AMIGA-map}}
\end{center}
\end{figure}

The cosmic ray spectrum presents a further feature at lower energy
named the knee ($\sim 4 \times 10^{15 }$ eV), outside the energy
region of the Auger Observatory. The knee has been interpreted
(Aglietta et al. (2004), Antoni et al. (2005), Kampert et al.
(2008)) as the spectrum region where galactic sources fail to
accelerate lighter elements to higher energies and only do so to
heavier elements. Still and as the energy increases, the number of
galactic sources capable of accelerating even the heavier elements
will diminish and therefore a transition from galactic to
extragalactic sources will take place. This transition is assumed
to occur either at the second knee (Nagano et al. (1984),
Abu-Zayyad et al. (2001), Pravdin et al. (2003), Abbasi et al.
(2004)) or at the ankle (Ave et al. (2001), Pravdin et al. (2003),
Abbasi et al. (2004)) and the way to help to identify it would be
a change in the cosmic ray composition. Within the Auger baseline
design described above, the surface array is fully efficient above
$ \sim 3 \times 10^{18}$ eV and in the hybrid mode this range is
extended down to $ \sim 10^{18 }$ eV which does not suffice to
cover the second knee - ankle region. For this purpose, Auger has
two enhancements: AMIGA (``Auger Muons and Infill for the Ground
Array'') (Etchegoyen et al. (2007), Platino et al. (2009),
Buchholtz et al. (2009), Medina-Tanco et al. (2009)) and HEAT
(``High Elevation Auger Telescopes'') (Klages et al. (2007),
Kleifges et al. (2009)). AMIGA is being deployed over a small area
of 23.5 km$^{2}$ (see Fig. \ref{AMIGA-map}) since the cosmic ray
flux abruptly increases with diminishing impinging energy. On the
other hand, the detectors have to be deployed at shorter distances
among each other in a denser array since lower energies imply
smaller air shower footprints on the ground. Also, telescopes need
to have a higher elevation field of view (FOV) since these showers
would develop earlier in the atmosphere.

\section{Spectrum, anisotropy, and composition}

\begin{figure}[!ht]
  \centering
  \includegraphics[width=.45\textwidth]{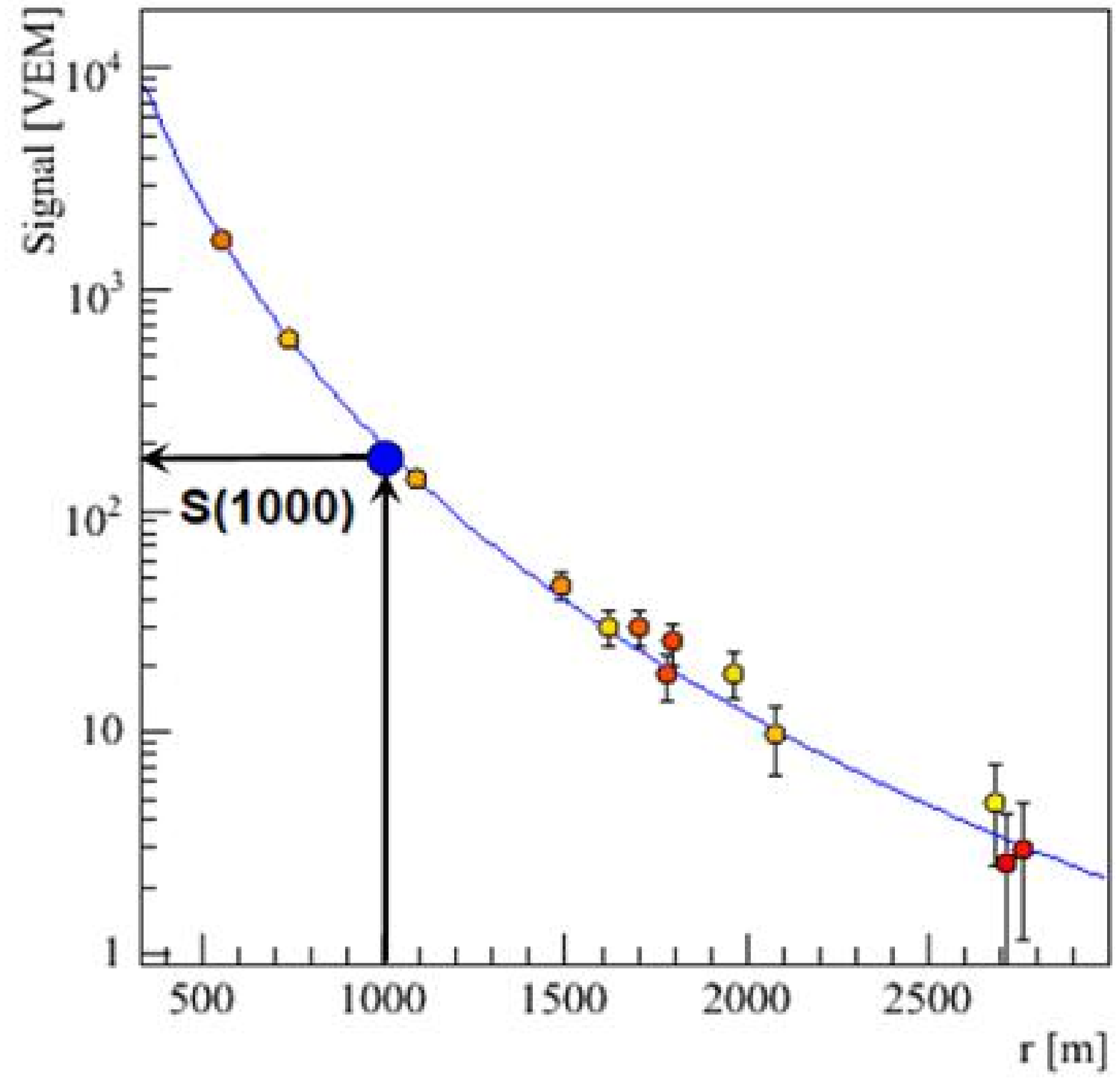}~\hfill
  \includegraphics[width=.45\textwidth]{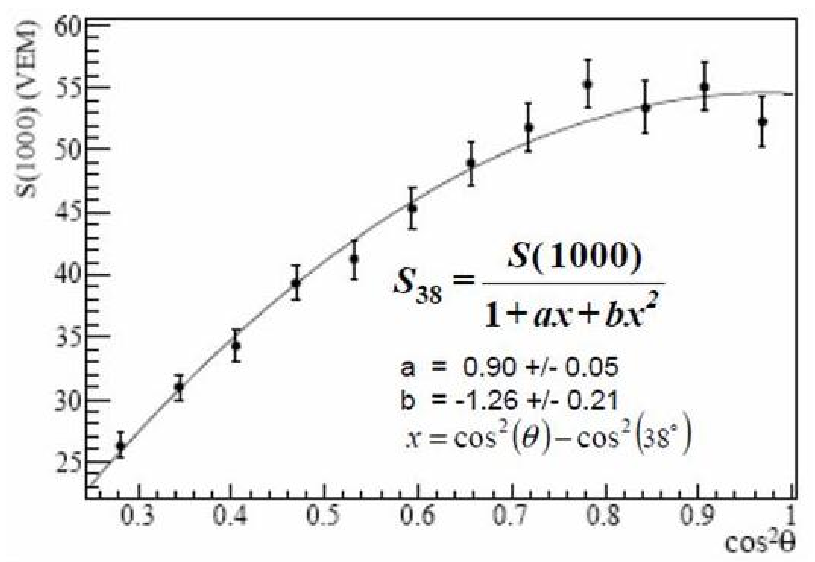}
  \caption{{\it (lhs)} Shower lateral profile; {\it (rhs)} Shower attenuation curve.}
  \label{fig:latp}
\end{figure}

\begin{figure}[!ht]
  \centering
  \includegraphics[width=.45\textwidth]{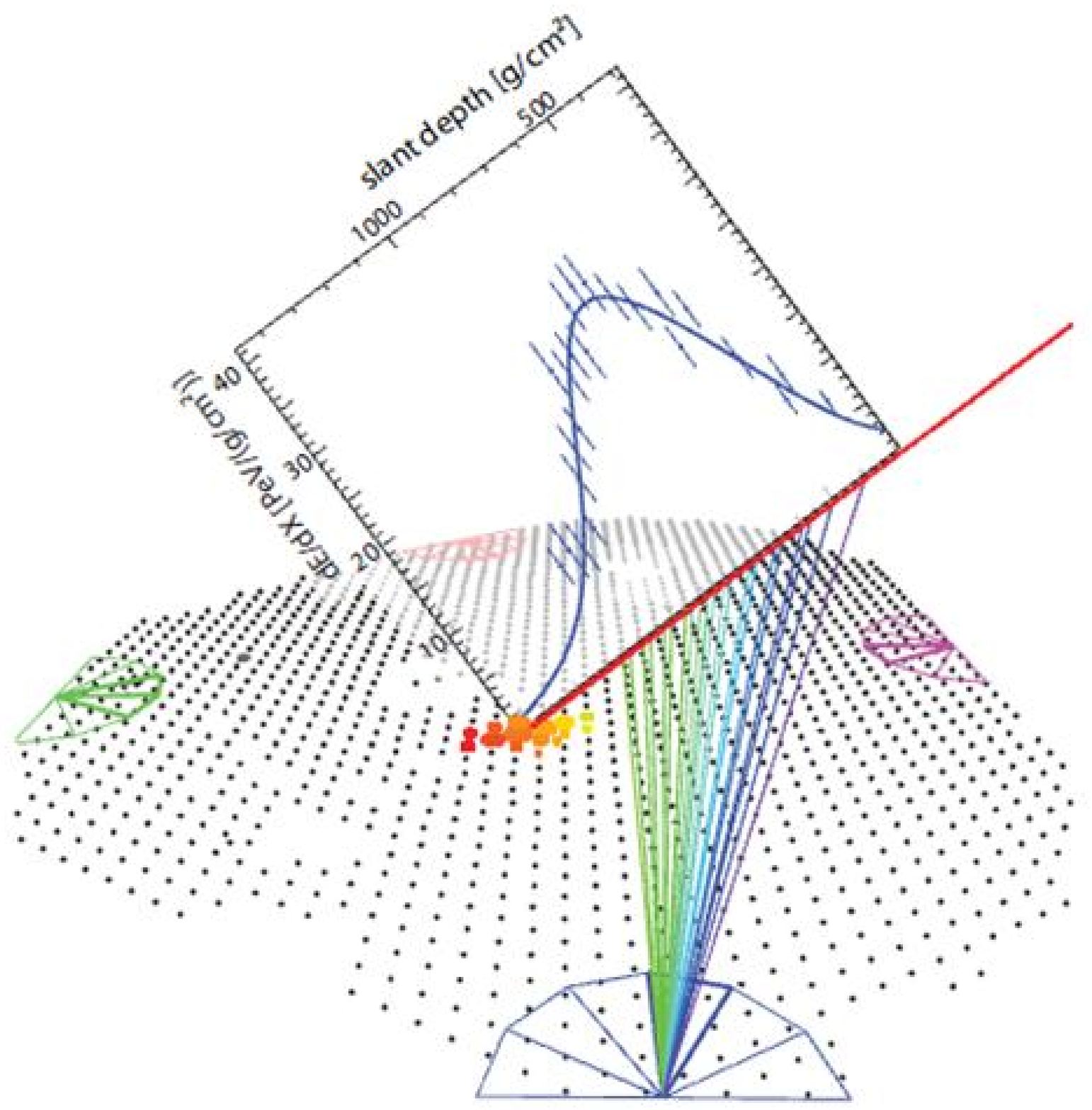}~\hfill
  \includegraphics[width=.45\textwidth]{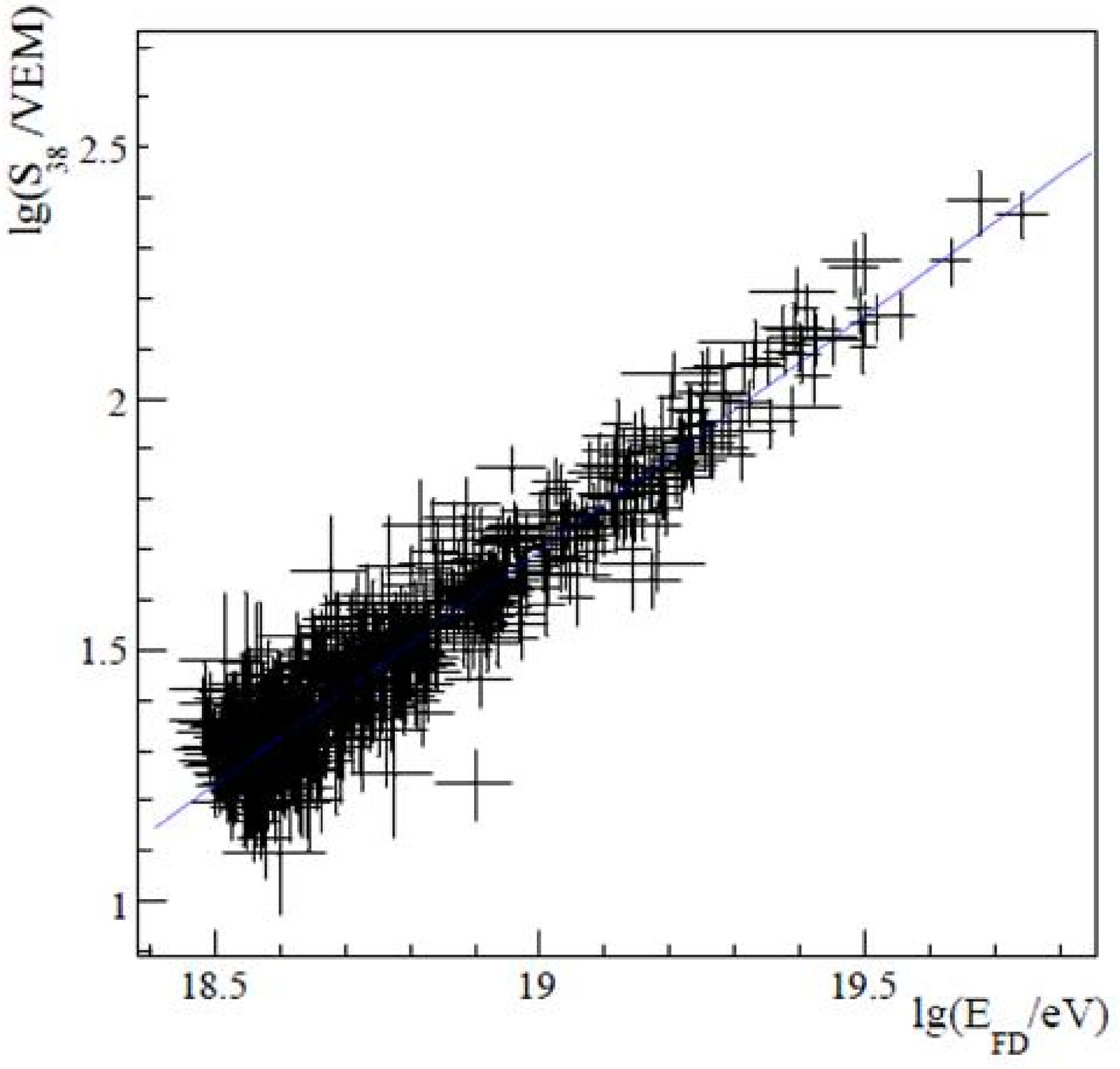}
  \caption{{\it (lhs)} Shower longitudinal profile; {\it (rhs)} Energy calibration curve.}
  \label{fig:lonp}
\end{figure}

The Auger Observatory as already mentioned has a hybrid detection
system composed of SD and FD systems. The SD array has a flat
exposure determined by the array geometry, a 100\% duty cycle and
a $\geq 1^\circ$ arrival directions uncertainty. Its energy
assignment depends on simulations which assume a hadronic model
and a primary composition. A hybrid measurement is performed when
an FD system and at least a SD station are triggered. This
combined time measurement improves the arrival direction
reconstruction to $\geq 0.2^\circ$. The energy assignment is
performed by integration of the longitudinal profile rendering a
calibration quite independent from both model and composition.
Still, the hybrid (or FD) exposure increases with energy and it
only has a $\sim 10\%$ duty cycle. The hybrid energy calibration
has several systematic uncertainties (fluorescence yield, absolute
calibration, atmospheric attenuation, and reconstruction method)
which have been estimated to give an overall energy resolution of
$\Delta E / E \sim 22 \%$. The Auger Collaboration calibrates the
SD array with hybrid events in order to avoid any dependence on
simulations. The procedure follows four steps: i) the lateral
profile is fitted with a carefully chosen lateral distribution
function and the signal, S(1000), at a core distance of 1000 m is
extracted, see Fig.\ref{fig:latp}.lhs, ii) the difference in
signal attenuation due to different atmospheric depths traversed
is experimentally corrected by renormalizing the signal to the
signal the shower would have produced with an arrival angle of
$38^\circ$, see Fig.\ref{fig:latp}.rhs, iii) the shower energy is
evaluated from the longitudinal profile measured by the FD system,
see Fig.\ref{fig:lonp}.lhs, and iv) the calibration curve is
obtained by plotting lg(S$_{38}$) vs lg(E$_{FD}$), see
Fig.\ref{fig:lonp}.rhs.

\begin{figure}[!ht]
\begin{center}
\includegraphics[width=8 cm]{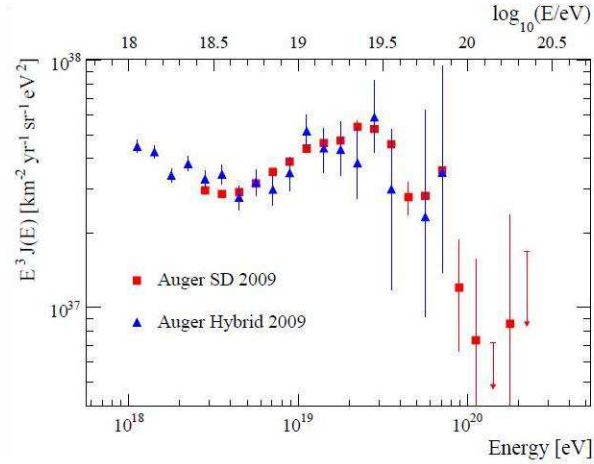}
\caption{Auger spectrum \label{fig:spec}}
\end{center}
\end{figure}

With the above mentioned calibration, the spectrum of cosmic rays
can be obtained with the large number of events measured by the
100\% duty cycle SD array. But the SD array is fully efficient for
E $\geq 3 \times 10^{18}$ eV while the hybrid system from E $\geq
10^{18}$ eV and therefore the total spectrum can be measured from
this latter energy onwards. Details on the event selection and
applied quality cuts can be seen in (Abraham et al. (2009a)). The
measured flux versus primary energy is displayed in
Fig.\ref{fig:spec} for both hybrid and SD detections (the flux is
multiplied by E$^{3}$ in order to highlight the ankle and the
cutoff features). It is clearly seen that both data sets give
quite consistent results, no systematic deviations in neither flux
nor energy. The ankle is seen at $\sim 10^{18.6}$ eV and the
GZK-cutoff at $\sim 10^{19.5}$ eV (Abraham et al. (2009a)). Both
are clear and notorious features, the spectrum is reaching its
end.

\begin{figure}[!ht]
  \centering
  \includegraphics[width= 8 cm]{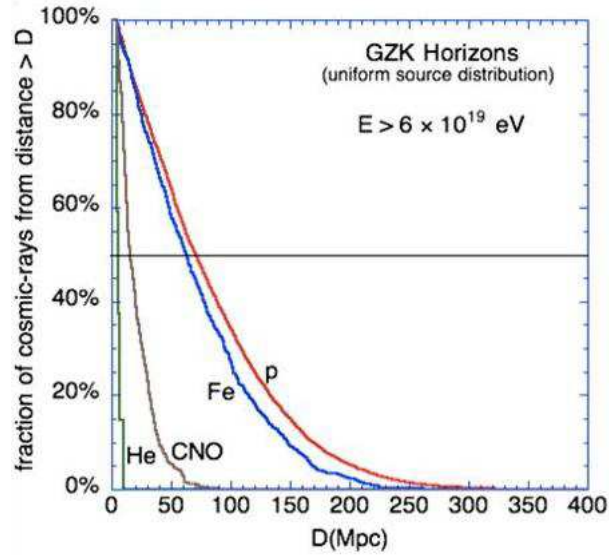}
  \caption{ GZK horizons, only $Fe$ and $p$ may survive $D \geq 100$ Mpc,
other nuclei will rapidly disappear for $D \geq 20$ Mpc.}
  \label{fig:hor}
\end{figure}

The cutoff is consistent with the predicted energy loss by the
interaction of extragalactic cosmic rays with the microwave
background radiation (Greisen (1966), Zatsepin et al. (1966)).
Also, this process will favor proton or iron primaries from the
nearby universe (see Fig.\ref{fig:hor}).

Anisotropy in the arrival directions of above the cutoff ($\sim
10^{18.6}$ eV) may appear since the GZK energy loss process
confines the possible cosmic rays sources to only a few in the
nearby universe. This feature opens the possibility to detect such
sources for their posterior study opening a field of novel
astronomy, charged-particles astronomy. A first anisotropy search
was published (Abraham et al. (2008b)) for cosmic rays with $E\geq
55$ EeV and a comparison was performed with the positions of AGN
galaxies listed in the Véron-Cetty and Véron (VCV) catalogue. The
search was performed for AGN within 75 Mpc with circles of radius
3.1$^{\circ}$ around the AGN. The current data set (Hague et al.
(2009)) is composed of 58 events with 26 events correlating with
nearby AGNs. The probability that an isotropic distribution would
by chance mimic these numbers is at the level of 1\%. Note also,
that the VCV catalogue is incomplete and inhomogeneous,
particularly near the galactic plane, and if events were removed
with galactic latitude $|b|<12^{\circ}$ the correlation becomes 25
out of 45 thus increasing the departure from an isotropic
distribution.

An interesting feature is that the region with the largest over
density is close to Cen A (an AGN only $\sim 3.5 Mpc$ away) from
where 12 events are detected within 18$^{\circ}$ while only 2.7
are expected from isotropy, i.e. a 2\% chance probability.

\begin{figure}[!ht]
  \centering
  \includegraphics[width=.45\textwidth]{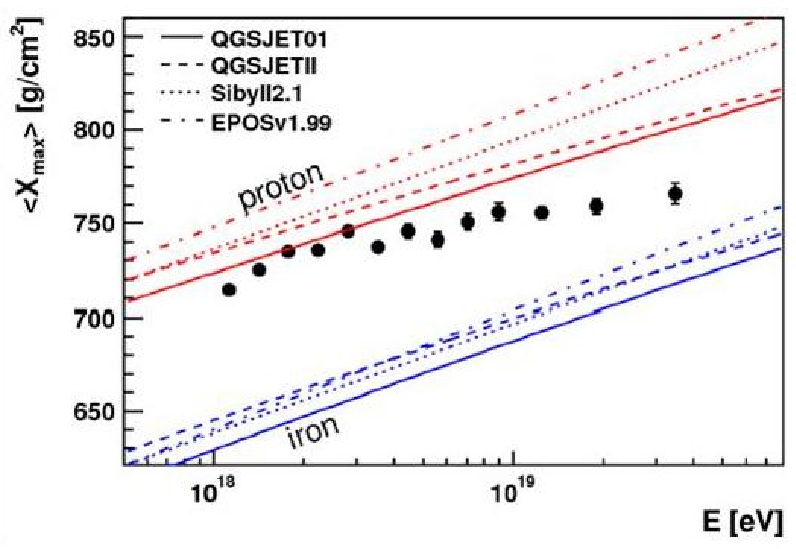}~\hfill
  \includegraphics[width=.45\textwidth]{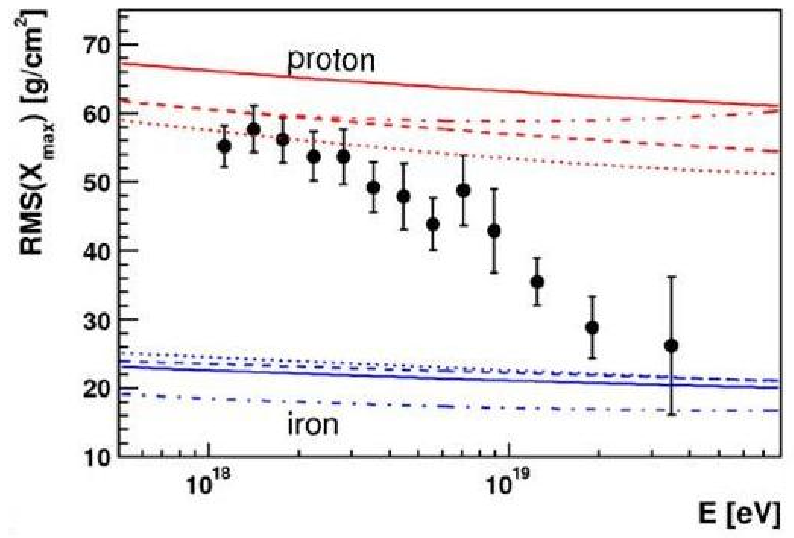}
  \caption{{\it (lhs)} Elongation rate (average atmospheric depth of
  shower maximum versus primary energy); {\it (rhs)} Average RMS(X$_{max}$) versus primary energy.}
  \label{fig:xmax}
\end{figure}

The chemical composition of primaries is of fundamental and unique
importance to understand cosmic ray sources. Actually, efforts are
concentrated in studying composition changes rather than absolute
composition since the latter would be quite dependent on hadronic
interaction models (see limits in Fig. \ref{fig:xmax}) and would
rely on comparisons to air shower simulations. A usual way to
assess composition is by measuring the average depths of shower
maximum, X$_{max}$ (and their fluctuations) which are obtained
from the air shower longitudinal profiles observed by the
fluorescence telescopes (Bellido et al. (2009), Abraham et al.
(2009b)).

Fig. \ref{fig:xmax} displays the composition studies performed
with Auger hybrid data. There is an indication both in the
elongation rate and in X$_{max}$ fluctuations that primaries
become heavier with energy (assuming the hadronic interaction does
not significantly change in the energy range of interest). The
elongation rate was satisfactory fitted (Abraham et al. (2009b))
with two slopes and the break was found at $\sim 10^{18.25}$ eV, a
factor of $\sim$ 2 below the reported ankle. These results
disagree with those found by the combined data from a hybrid
experiment (HiRes/MIA), and the HiRes stereo fluorescence
experiment, which overlap by only $\sim$ 1/4 of a decade in energy
(Sokolsky et al. (2005)). HiRes observes that the composition gets
lighter and remains so starting at energies close to the second
knee.

\section{Auger Observatory Enhancements}

\begin{figure}[!ht]
  \centering
  \includegraphics[width=.50\textwidth]{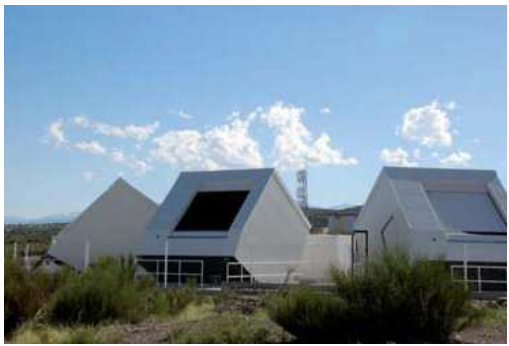}~\hfill
  \includegraphics[width=.40\textwidth]{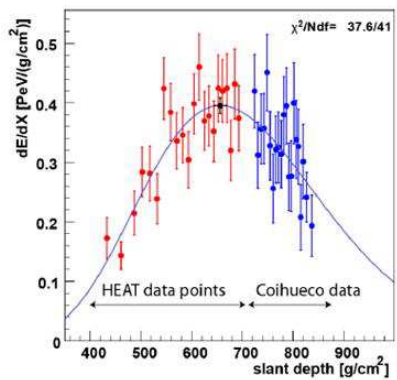}
  \caption{(Kleifges et al. (2009)){\it (lhs)}  The three buildings of the HEAT telescopes at Cerro Coihueco;
  {\it (rhs)} Longitudinal shower profile where both HEAT and Coihueco telescopes are
  needed in order to reconstruct the profile, event energy $(2.0 \pm 0.2) 10^{17}$ eV.}
  \label{fig:heat}\end{figure}

As already mentioned the cosmic ray flux shows three high-energy
features, the second knee, the ankle and the GZK cutoff. There are
currently no doubts about the existence of these three traits, but
the understanding of the first two is yet under scrutiny with
conflicting results as shown in the previous section. To fully
cover this energy region with a single Observatory the Auger
Collaboration is building two enhancements, HEAT and AMIGA (Klages
et al. (2007), Kleifges et al. (2009), Medina et al. (2006),
Etchegoyen (2007), Supanitsky et al. (2008), Platino et al.
(2009), Buchholtz et al. (2009), Medina-Tanco et al.(2009)). These
enhancements will have an enlarged energy range down to 10$^{17}$
eV and therefore will allow for a good comparison with
Kascade-Grande results (Haungs et al. (2009)) whose primary
objectives are to study the cosmic ray primary composition and the
hadronic interactions in the energy range $10^{16} - 10^{18}$ eV.
Fig. \ref{fig:heat} shows the HEAT telescopes enclosures and a
reconstructed longitudinal profile of a low energy event.

AMIGA data acquisition has started with the SD 750 m infilled area
and preliminary analyzes have already been performed with a data
set restricted to events with zenith angle $\theta \leq
60^{\circ}$ and well contained within the infill (i.e. with the
six SDs of the hexagon enclosing the highest signal SD in its
center active). Preliminary tests were performed reconstructing
events with and without the SDs from the infilled area in order to
probe the main array reconstruction uncertainties (Platino et al.
(2009)). Emphasis was also focused on the preliminary energy
calibration by relating the ground signal at 600 m from the core,
S(600), with the energy measured by FD (note that the ground
signal is taken at 600 m rather than 1000 m since the array
spacings are different) (Platino et al. (2009)).

\begin{figure}[!ht]
  \centering
  \includegraphics[width=.45\textwidth]{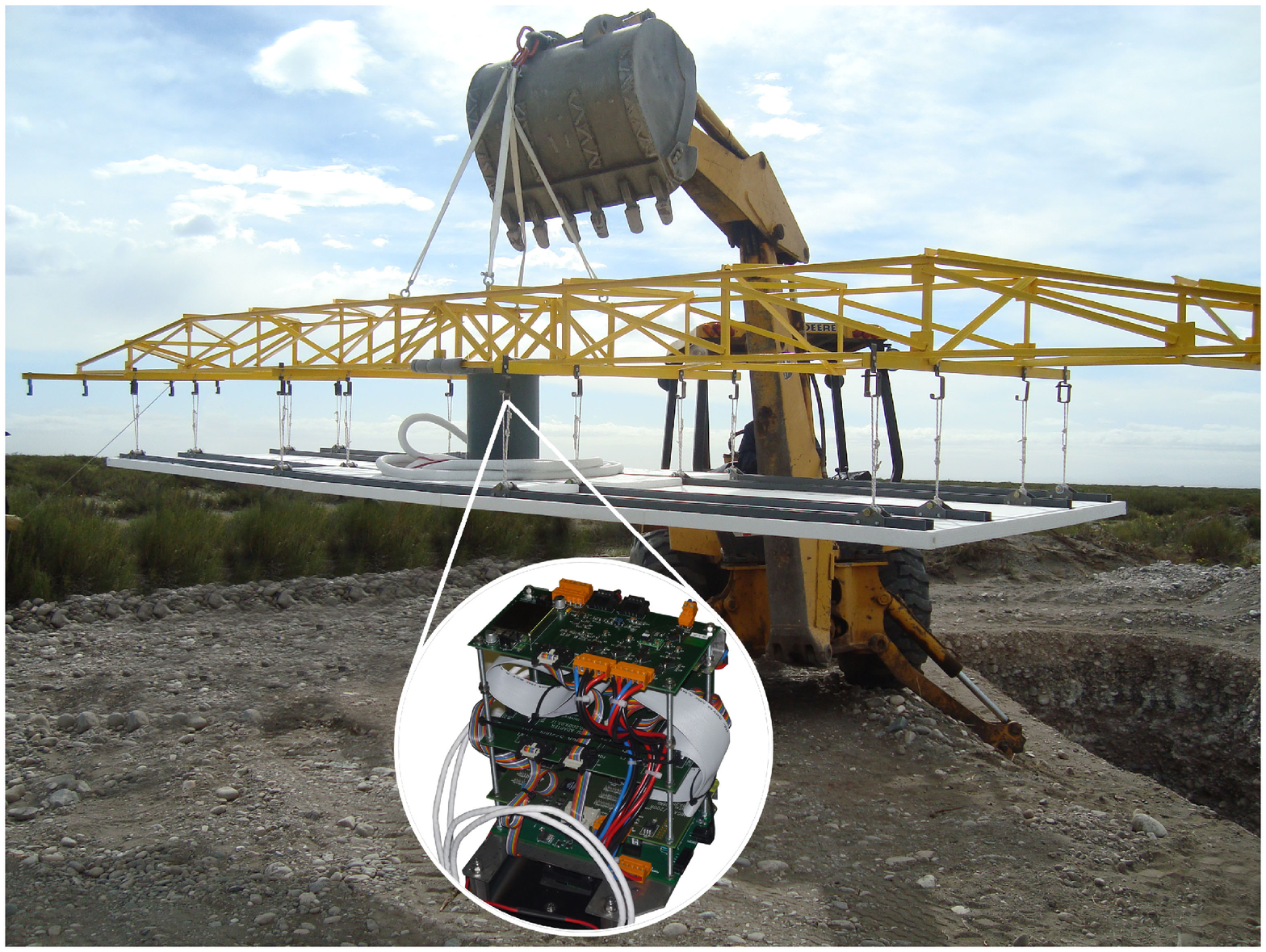}~\hfill
  \includegraphics[width=.45\textwidth]{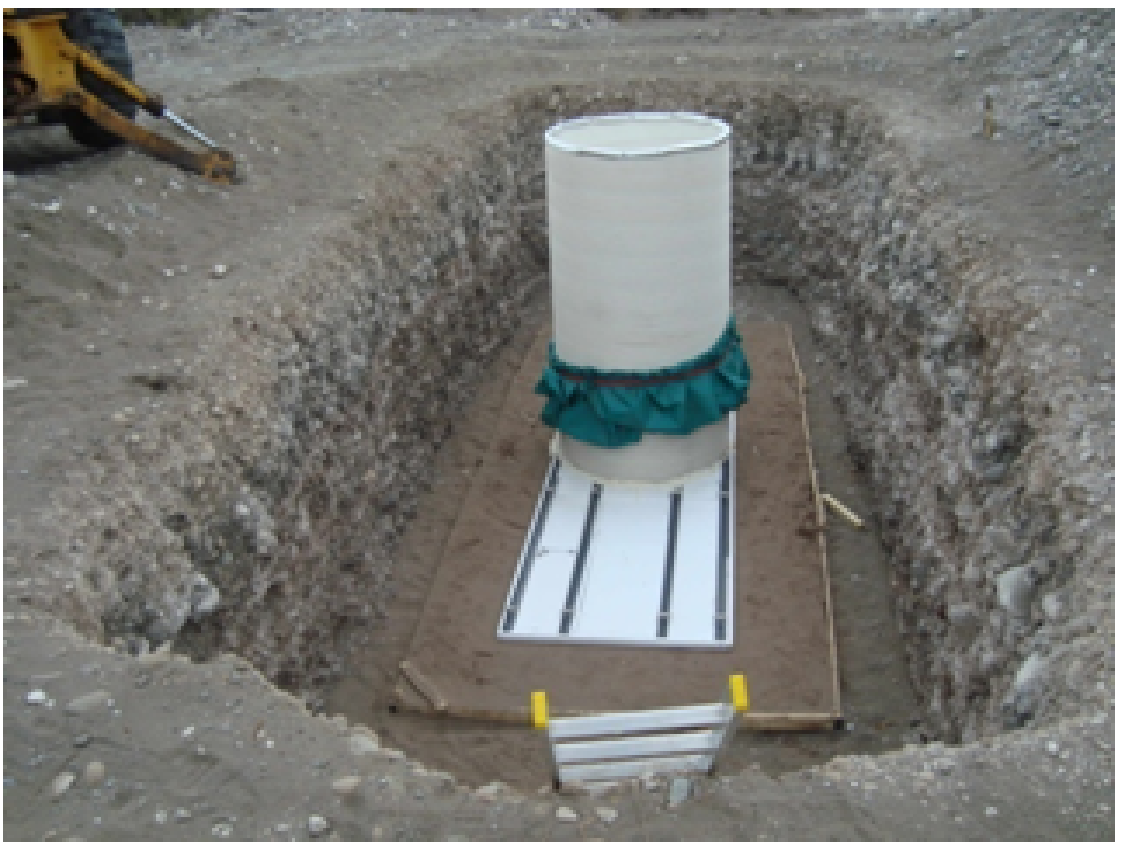}
  \caption{{\it (lhs)} First module buried at the Observatory site, the insert shows the electronics inside its
  enclosure;
  {\it (rhs)} Muon counter placed over sand bed with service pipe installed, about to be buried.}
  \label{fig:counter}
\end{figure}

Apart from the extended energy range, the enhancements include
muon counters for composition analyses buried alongside each
surface detector of the infilled area. Each muon counter has an
area of 30 m$^{2}$ and it is made of scintillator strips with
glued optical fibers (doped with wave length shifters) in a groove
coated on top with a reflective foil. The strips are 1 cm thick
and 4.1 cm wide. The counters of the Unitary Cell(an hexagon with
7 detector pairs, one in each hexagon vertex and one in the
center) are composed of 4 modules each, 2 $\times$ 5 m$^{2}$ and 2
$\times$ 10 m$^{2}$ with 2 and 4 m long strips, respectively. Each
module has a 64 pixel Hamamatsu H8804MOD photomultiplier tube with
a 2 mm $\times$ 2 mm pixel size. The front end bandwidth is of 180
MHz, the electronics sampling is performed at 320 MSps (3.125 ns)
with a memory to store up to 6 ms of data. The total number of
independent electronic channels per counter is 256 due to the high
segmentation requirement in an effort to measure no more than a
single muon por segment per unit time. The first fully equipped 5
m$^{2}$ module has been designed, built, tested, and buried at the
Observatory site, see Fig.\ref{fig:counter}.

The extension of Auger to full efficiency down to 0.1 EeV with a
direct measure of both X$_{max}$ and shower muon contents will
open an unique experimental tool to study the cosmic ray spectrum
in the second knee - ankle region, where the transition from
galactic to extragalactic sources is assumed to occur. Data with
unprecedented precision will be available and in particular there
will be a triple hybrid data set (fluorescence, muon, and surface
detector detections) on which careful primary energy and
composition analyses could be performed.

In conclusion, the Auger Observatory has measured the high energy
cosmic ray spectrum and clearly identified the ankle and the
cutoff, also it found clear indications of anisotropy in the
arrival directions of cosmic rays with energies above 55 EeV.
Statistical composition analyses have been performed with depths
at shower maximum and their fluctuations. Enhancements are well
under way in order to study the transition region from galactic to
extragalactic sources with surface, telescope, and muon detectors
with unitary efficiencies unbiased in composition.

\begin{referencias}
% {\bf NOTA:} Las citas bibliográficas deberán seguir el estilo habitual
% de las publicaciones astronómicas (por ejemplo Astronomy \& Astrophysics).
% Trabajos de un solo autor se citarán: Cioran (1983); trabajos de dos autores:
% Arlt \& Marechal (1939); y trabajos de tres o más autores: Borges
% et al. (1934)

%Abu:01
\reference T. Abu-Zayyad et al., 2001, Astrophys. J 557, 686.

%Antoni:05
\reference Antoni et al., 2005, Astropart. Phys. 24, 1.

%Auger-Hague:09
\reference J.D. Hague for the Pierre Auger Collaboration, 2009,
Proc. 31st ICRC (Lodz, Poland), \#0143.

% Auger-EA:04
\reference The Auger Collaboration (J.Abraham et al.), 2004,
  Nucl. Inst. \& Meth. A532, 50, 95.

% Auger-SD:08
\reference The Auger Collaboration (J.Abraham et al.), 2008a,
Nucl. Inst. \& Meth. A586, 409, 420.

%Auger-LongAGN:08
\reference The Auger Collaboration (J. Abraham et al.), 2008b,
Astropart. Phys. 29, 188, 204.

% Auger-Schu:09
\reference The Auger Collaboration (J. Abraham et al.), submitted
to Physics Letters, (2009a).

%Auger-Unger:09
\reference The Auger Collaboration, submitted to Phys. Rev. Lett.,
(2009b).

%Ave:01
\reference M. Ave et al., 2001, Proc. 27$^{th}$ ICRC (Hamburg),
381.

%Auger-Bellido:09
\reference J. A. Bellido, for the Pierre Auger Collaboration,
2009, Proc. 31st ICRC (Lodz, Poland), \#0124.

%Buchholtz:09
\reference P.Buchholtz for the Pierre Auger Collaboration, 2009,
Proc. 31st ICRC (Lodz, Poland), \#0043

%Aglietta:04
\reference The EAS-TOP and MACRO Collaborations (M. Aglietta et
al.), 2004, Astropart. Phys. 20, 641.

%Etchegoyen:07
\reference A. Etchegoyen for the Pierre Auger Collaboration, 2007,
Proc. 30th ICRC (M\'erida-M\'exico), \#1307.

%Greisen:66
\reference Greisen K., 1966, Phys. Rev. Lett. 16, 748.

%Abbasi:04
\reference The HiRes Collaboration (R.U. Abbasi et al.), 2004,
Phys. Rev. Lett. 92, 151101.

%Kampert:08
\reference The Kascade Collaboration (K.H. Kampert et al), 2008,
arXiv:astro-ph/0405608.

%Kascade-Grande:09
\reference The Kascade-Grande Collaboration (A. Haungs et al.),
2009, Proc. 31st ICRC (Lodz, Poland), \#0401.

%Klages:07
\reference H. Klages for the Pierre Auger Collaboration, 2007,
Proc. 30th ICRC (M\'erida-M\'exico), \#0065.

%Kleifges:09
\reference M. Kleifges for the Pierre Auger Collaboration, 2009,
Proc. 31st ICRC (Lodz, Poland), \#0410.

%Medina:06
\reference M.C. Medina et. al., 2006, Nucl. Inst. and Meth. A566,
302, 311, and astro-ph0607115.

%Medina-Tanco:09
\reference G. Medina-Tanco for the Pierre Auger Collaboration,
2009, Proc. 31st ICRC (Lodz, Poland), \#0137

%Nagano:84
\reference M. Nagano et al., 1984, J. Phys. G 10, 1295.

%Platino:09
\reference M. Platino for the Pierre Auger Collaboration, 2009,
Proc. 31st ICRC (Lodz, Poland), \#0184.

%Pravdin:03
\reference M.I. Pravdin et al., 2003, Proc. 28$^{th}$ ICRC
 (Tuskuba), 389.

%HiRes-Composition:05
\reference P. Sokolsky, John Beltz, and the HiRes Collaboration,
2005, Proc. 29th ICRC (Pune, India), 101, 104.

%Supanitsky:08
\reference A.D. Supanitsky et al., 2008, Astroparticle Physics 29,
461, 7470.

%Zatsepin:66
\reference Zatsepin G. T., Kuzmin V. A., 1966, Sov. Phys. JETP
Lett. 4, 78.

\end{referencias}

\end{document}